\documentclass[times,twocolumn,final]{elsarticle}

\usepackage{medima}
\usepackage{framed,multirow}

\usepackage{amssymb}
\usepackage{latexsym}

\usepackage{url}
\usepackage{xcolor}

\usepackage{hyperref}

\definecolor{newcolor}{rgb}{.8,.349,.1}

\journal{preprint}
\graphicspath{{./}}
\begin{document}

\verso{Gillespie \textit{et~al.}}

\begin{frontmatter}

\title{Deep learning in magnetic resonance prostate segmentation: A review and a new perspective}

\author[1]{David \snm{Gillespie}\fnref{fn1}}
\author[1]{Connah \snm{Kendrick}}
\author[2]{Ian \snm{Boon}} 
\author[3]{Cheng \snm{Boon}}
\author[4]{Tim \snm{Rattay}}
\author[1]{Moi Hoon \snm{Yap}\corref{cor1}}
\cortext[cor1]{Corresponding author: 
	Tel.: +44 161 247 1503;  
}
\ead{M.Yap@mmu.ac.uk}

\address[1]{Manchester Metropolitan University, John Dalton Building, Chester Street, Manchester M1 5GD, UK}
\address[2]{NHS, The Leeds Teaching Hospital, Great George St, Leeds, LS1 3EX, UK}
\address[3]{NHS, The Clatterbridge Cancer Center, Lower Lane, Liverpool L9 7AL, UK}
\address[4]{Leicester University,Leicester Cancer Research Centre, University Road, Leicester, LE1 7RH, UK}


\begin{abstract}

 Prostate radiotherapy is a well established curative oncology modality, which in future will use Magnetic Resonance Imaging (MRI)-based radiotherapy for daily adaptive radiotherapy target definition. However the time needed to delineate the prostate from MRI data accurately is a time consuming process. Deep learning has been identified as a potential new technology for the delivery of precision radiotherapy in prostate cancer, where accurate prostate segmentation helps in cancer detection and therapy. However, the trained models can be limited in their application to clinical setting due to different acquisition protocols, limited publicly available datasets, where the size of the datasets are relatively small. Therefore, to explore the field of prostate segmentation and to discover a generalisable solution, we review the state-of-the-art deep learning algorithms in MR prostate segmentation; provide insights to the field by discussing their limitations and strengths; and propose an optimised 2D U-Net for MR prostate segmentation.  We evaluate the performance on four publicly available datasets using Dice Similarity Coefficient (DSC) as performance metric. Our experiments include within dataset evaluation and cross-dataset evaluation. The best result is achieved by composite evaluation (DSC of 0.9427 on Decathlon test set) and the poorest result is achieved by cross-dataset evaluation (DSC of 0.5892, Prostate X training set, Promise 12 testing set). We outline the challenges and provide recommendations for future work. Our research provides a new perspective to MR prostate segmentation and more importantly, we provide standardised experiment settings for researchers to evaluate their algorithms. Our code is available at https://github.com/AIEMMU/MRI\_Prostate.
\end{abstract}


\end{frontmatter}


\section{Introduction}
\label{sec1}
Following an initial diagnosis of cancer, many patients are referred to specialist centres for radiotherapy. Radiotherapy is an important part of the potentially curative treatment of both breast and prostate cancer, where it  is used in 4 out of 10 of all cancer curative treatment modalities \cite{cullen2019recommendations}. The number of radiotherapy courses in Europe was estimated at 1.7M in 2012. The largest number of radiotherapy courses was for breast cancers (396,891) and second largest was for prostate cancer (243,669) \citep{borras2016many}. The projected incidence of cancer continues to rise with prostate cancer being the most commonly diagnosed cancer in the United Kingdom \cite{prostatecanceruk}. The evidence for both curative \cite{dearnaley2016conventional} and palliative radiotherapy in prostate cancer is well established \cite{parker2018radiotherapy}.

The future of prostate radiotherapy will be the development of  MR-only based radiotherapy \cite{pathmanathan2018magnetic}. In MRI-only prostate radiotherapy pathway, the significant benefit would be MRI allows better soft tissue definition, which means precision radiotherapy can be further improved. Precision radiotherapy allows potentially increased in cure rates and further avoidance of normal tissue toxicities \cite{pathmanathan2018magnetic}. Prostate cancer patients have a high cure rate, and therefore, any reduction of long term side effect would be advantageous. The challenge of MR-only radiotherapy is it requires daily adaptive radiotherapy contouring for delivery. herefore this paper will only address the use of MRI data for delineating the prostate volume. 

At present, clinical oncologists manually contour or delineate the primary cancer volume as part of the target volume definition process of the radiotherapy workflow. In contouring prostate cancer in the curative setting, firstly clinicians will delineate the primary cancer target volume. This will be the primary prostate volume, and in high risks, prostate cancer clinicians may potentially include seminal vesicles structures and suspected nodal volumes. Next, clinicians will outline adjacent Organ At Risks (OARs) including the rectum, bowels, bladder, and femoral heads which are the normal structures to avoid radiotherapy doses. However, this contouring process has been identified as the weakest link with wide-ranging variation amongst clinicians \cite{vinod2016uncertainties}. This is where auto-segmentation using deep learning algorithms can be helpful. 

Artificial Intelligence (AI) has been identified by the UK NHS and the Royal College of Radiologists as a major technological advancement to radiology and oncology \cite{topol2019topol}, for the use of deep learning image segmentation. Image segmentation is the process of segmenting an image into various classes based on their pixel values. It is widely used in various fields such as autonomous driving, visual tracking, but it is still one of the most challenging aspects of analysing medical images. The task of segmenting organs or lesions from the background of CT or MRI has improved recently with the development of Convolutional Neural Networks (CNNs) and the use of Fully Convolution Networks (FCNs) \cite{long2015fully}. The introduction of the U-Net \cite{ronneberger2015u}, Residual Convolution Blocks \cite{alom2018recurrent} and attention mechanisms \cite{oktay2018attention} has significantly improved performance and detections over earlier systems based on edge detection algorithms. The promising nature of deep learning segmentation approaches has put them as a primary option and solution for health care professionals. 

However, while literature discusses the potentials of deep learning image segmentation approaches, there are few reviews on the pitfalls of architectures or the discussion regarding data acquisition. While new models can achieve state-of-the-art benchmarks on publicly available datasets, these scores are rarely reflected when used on real-world data acquired from clinical institutions. The lack of large prostate segmentation datasets, or reproducible results lead to the state-of-the-art segmentation models failing to generalise to real-world scenarios.

The aim of this paper was to review state-of-the-art models used in prostate MRI segmentation, to discuss their limitations and strengths, and to conduct a study comparing performance of an optimised 2D U-Net to segment the prostate in four publicly available prostate MRI datasets. We also discuss the potential of creating a protocol for the evaluation and testing of future algorithms, to allow for reproducibility and to be explored by our fellow peers.  

\section{Review on Existing Methods}
\label{sec2}
Early attempts to automate the process of delineating the prostate relied on edge or region based models, which incorporated prior knowledge or used atlas-based segmentation \cite{sharp2014vision,iglesias2015multi,klein2008automatic}. They use an approach that matches manually selected regions on a target image to selected reference images, where the ground truth reference segmentation mask is mapped back to the target image. However, these approaches are heavily reliant on the atlas selection strategy as well as how well the region was selected \cite{schipaanboord2019evaluation,peressutti2016tu}. To improve upon atlas-based approaches, deep learning methods have been applied to medical imaging segmentation tasks resulting in significant improvements, especially with the recent developments of CNNs, FCNs \cite{long2015fully} and the U-Net \cite{ronneberger2015u}. Deep learning approaches can learn a rich set of features to perform pixel-wise image segmentation or classification \cite{ronneberger2015u,guo2015deformable} offering a computation advantage over atlas-based methods, as they can inference directly on new data. 

A CNN is a neural network that learns by feeding input into a network, that consists of layers (convolutions, maxpools, relu), where the output from one layer is fed into the next layer. Through this process, the network learns to recognise patterns, and abstractions to be able to classify input data correctly \cite{long2015fully,bibault2016big}. The use of CNNs has achieved outstanding results for automatically detecting prostate cancer in multi panoramic MRI \cite{yang2017co, wang2018automated}, as well as patch-based ensembling methods for patch-based segmentation \cite{jia2018atlas}.

 A FCN \cite{long2015fully} is similar to a CNN, however, it does not contain any dense layers that are common in CNN, and the last layer is replaced with a simple convolutional layer with a stride of 1 and a kernel size of 1. This allows the network to output a feature map that, contains pixel-wise predictions for each class from the full image.  Ronneberger et al. \cite{ronneberger2015u} expanded upon the FCN with the development of the U-Net. It uses a CNN as a downward sampling path, with an up-sampling operation, increasing the resolution of the output feature maps. It also uses skip connections, to provide more information to the feature maps that are placed within the up-sampling operations \cite{ronneberger2015u}. The U-Net is one of the most popular networks for image segmentation on medical images \cite{zhou2018unet++, cciccek20163d,alom2018recurrent,isensee2018nnu,oktay2018attention,dong2017automatic,ibtehaz2020multiresunet,alom2019recurrent}. 

Clark et al. \cite{clark2017fully, clark2017fullya} was one of the first to deploy a U-Net to delineate the prostate. They use a trained standard VGG network \cite{simonyan2014very}, to detect slices in Diffusion-Weighted MRI (DWI) that contained a part of the prostate. They then used a U-Net to segment the whole prostate gland and transitional zone, before using a U-Net to segment the Prostate. Using this simple approach, they were able to achieve a DSC of 0.93 for the prostate on an in house dataset. However, when they validated on T2-weighted (T2W) MRI data from the Promise 12 challenge \cite{litjens2014evaluation}, they were only able to achieve a DSC of 0.86. Zhu et al. \cite{zhu2017deeply} proposed a deeply supervised U-Net architecture for prostate segmentation. It utilises residual blocks to reduce the number of parameters compared to the original U-Net \cite{zhu2017deeply} and uses a deep supervision strategy. The deep supervision strategy is to supervise the hidden layers within the network and propagate it to the lower levels of the network. Using more than one supervision within the network leads to enhancing the feature maps of the network, thus better segmentation \cite{zhu2019fully}. Zhu et al. \cite{zhu2017deeply} were able to out perform the original U-Net to segment the prostate, and scored a dice coefficient score of 0.885 on their own internal dataset.  

Tian et al. \cite{tian2018psnet} developed the PSNET from the FCN \cite{long2015fully}. The network was fine-tuned on an in-house dataset consisting of 42 T2W MRI volumes and using patients from two open-source prostate segmentation datasets, the PROMISE12 challenge \cite{litjens2014evaluation} and the ISBI2013 \cite{bloch2015nci} dataset. Using a five-fold cross-validation procedure using 112 out of 140 volumes for training, they achieved a validation DSC of 0.85. Karimi et al. \cite{karimi2018prostate} followed a similar approach, they used a smaller FCN consisting of 3 layers, and used a much smaller dataset consisting of 49 T2W axial MRI images and with 26 MRI images from the PROMISE12 challenge \cite{litjens2014evaluation} using patients that did not have an endorectal coil. Using a cross-validation scheme, Karimi et al. \cite{karimi2018prostate} was able to achieve a 0.88 DSC on their validation set. Both networks were able to achieve high DSCs on their respective validation datasets, demonstrating the potential of deep learning for MRI prostate image segmentation. 

The U-Net architecture has since been adapted to be able to segment prostates from 3D MRI volumes instead of MRI slices by Milletari et al. \cite{milletari2016v}. They trained their network on the Promise12 challenge dataset \cite{litjens2014evaluation}, where they show their network performed significantly better than other networks at the time. They also proposed a loss function based on the Dice coefficient. The use of this loss function and data augmentation allowed Milletari et al. \cite{milletari2016v} to achieve an even higher DSC than their implementation trained using a multi logistic loss function \cite{milletari2016v}. Yu et al. \cite{yu2017volumetric} improved upon this, by adding residual connections to the network into the 3D segmentation network, which helped to improve the prostate segmentation. It also used a sum operation instead of the concatenation operation features into the up-sampling layer, turning it into a ResNet-U-Net hybrid.

To improve upon the accuracy of prostate segmentation, Zhu et al. \cite{zhu2019fully} proposed the use of a cascaded U-Net. Where the output from the first network, segmented the whole prostate gland, and the segmented gland was fed into the second network to segment the peripheral zone. Using this method, they were able to achieve a mean DSC of 0.92 for segmenting the whole prostate, and 79.7 for the peripheral zone on their own dataset. Using a similar strategy as Zhu et al. \cite{zhu2017deeply}, Wang et al. \cite{wang2019deeply} proposed a deep supervision strategy for a 3D U-Net and group dilated convolutions for segmentation of the prostate gland. The deep learning strategy was proposed to prevent exploding or vanishing gradients when training deep models, but also to force the hidden layers to favour highly discriminate features. Wang et al. \cite{wang2019deeply} also used group dilated convolutions to extract more global contextual information, as they increase the receptive field of the network, without adding extra parameters and memory cost to the network. They also used a combined loss function consisting of cross-entropy loss and cosine, which measures the similarity and dissimilarity between segmentation and manual contours. These techniques improved the segmentation accuracy of their network, achieving a DSC of 0.86 on their internal dataset, and a score of 0.88 on the Promise-12 challenge dataset \cite{litjens2014evaluation}.

Liu et al. \cite{liu2019automatic} proposed a network to segment the prostate zones using FCN with a feature pyramid attention mechanism. They used a ResNet50 as the backbone of their network, with a feature pyramid network to capture features at multiple scales, and a simple decoder. The feature pyramid attention can capture strong semantic information at multiple scales, which are combined to generate a high-resolution feature representation. This helps to improve the segmentation results of Lui et al. \cite{liu2019automatic}, over the original U-Net \cite{ronneberger2015u}, where it achieved a DSC of 0.74 and 0.86 on the prostatic transitional zone and peripheral zone from 63 patients from the Promise 12 dataset \cite{litjens2014evaluation}. They were able also to achieve a DSC of 0.74 and 0.79 of the prostatic transitional zone and peripheral zone on an external dataset.

Zabihollahya et al. \cite{zabihollahy2019automated} proposed a U-Net network architecture that would segment the prostate zones on T2W and apparent diffusion coefficient (ADC) map prostate MRI images separately. They trained 2 networks per modality, one to segment the whole prostate gland and another to segment the central gland. They then use a post-processing process to create the contours of the whole prostate, central gland and the peripheral zone on both of the modalities \cite{zabihollahy2019automated}. Using this methodology, they were able to achieve on their own internal dataset a mean DSC of 0.92, 0.91 and 86.22 for the whole gland, central gland and peripheral zone respectively on the T2W MRI images. They were also able to achieve similar scores on the apparent diffusion coefficient MRI images with 0.89, 0.86 and 0.83 for the whole gland, central gland and peripheral zone \cite{zabihollahy2019automated}. 

 Nie and Shen \cite{nie2019semantic} proposed a semantic guided strategy to learn discriminative features, to improve delineation of the prostate. They also proposed a soft contour constraint mechanism to model the blurry boundary detections, to improve the segmentation of the prostate. They trained their network on 50 prostate cancer patients, with ground truths annotated manually of the rectum, prostate and bladder. Using a 5 k-fold validation scheme, they obtained a DSC of 0.975, 0.932, and 0.918 for the bladder, prostate and rectum respectively. Their approach outperformed state of the art techniques \cite{nie2019semantic} achieved a DSC of 0.92 on the Promise12 dataset \cite{litjens2014evaluation} and a score of 0.89 on the hidden test set of the Promise12 challenge. Zhu et al. \cite{zhu2019boundary} proposed a boundary-weighted segmentation loss, to increase the sensitivity of the network to boundary segmentation. They also introduced a boundary weighted transfer learning approach was exploited to overcome the challenge of small datasets that are available for training. Using this approach, and leveraging several datasets, they were able to achieve the state of the art results on the Promise12 Challenge test dataset \cite{litjens2014evaluation}, achieving a score of 0.8958.
 
 \subsection{Generalisation}
For prostate delineating, MRI scans normally have a third dimension. This third dimension allows for valuable information to be extracted from the Z axis, which makes a 3D U-Net the most appropriate choice \cite{isensee2018nnu}. However, this limits training data, and uses alot more memory for computation versus a 2D U-net. \cite{isensee2018nnu} proposed a novel approach, that condenses domain knowledge, and chooses the best nnU-Net configuration (2D or 3D) for a given dataset. This approach has managed to achieve state of the art results in 19 competitions at the time of its release \cite{isensee2019automated}. They trained a 2D and a 3D variant nnU-net with a 5 fold cross-validation scheme, with no manual intervention for both the Promise12 challenge \cite{litjens2014evaluation} and the Medical Decathlon dataset \cite{simpson2019large}. Using an ensembling approach with the output from their 3D and 2D nnU-Net, they managed to achieve a DSC of 0.8965 on the Promise12 Challenge \cite{litjens2014evaluation} and a 0.77 and 0.90 DSC for the peripheral zone and transitional zone on the Decathlon dataset \cite{simpson2019large}. In their paper, they also demonstrate that common modifications to U-Nets to improve segmentation results, do not always necessarily result in a good performance or are superior to a properly tuned baseline method \cite{isensee2019automated}. 

One of the main limitations of the above papers is their focus on single site prostate MRI vendors for network training to improve their performance. This allows for the model to perform well on their own data, but it does not take into consideration how it may perform on data from another site, or MRI vendor. This is a large issue for prostate segmentation, as there are differences between MRI scanner vendors and protocols \cite{rundo2019use,zavala2020segmentation,liu2020ms}. Rundo et al. \cite{rundo2019use} attempted to solve this issue by using a channel-wise feature recalibration to improve the segmentation results. They also incorporate squeeze and excitation blocks into their network \cite{rundo2019use}. They compared their network against U-Net \cite{ronneberger2015u}, Pix to Pix \cite{isola2017image}, MS-D net and a continuous max flow model \cite{qiu2014dual}. They performed well when their network was trained on all of the dataset's, but less so when trained only on one or two of the datasets \cite{rundo2019use}. However, their network was still robust, and their DSC score were very similar to the other networks it tested against. 

Another attempt to generalise to multi-site MRI images was performed by Zavala Romero et al. \cite{zavala2020segmentation} who implemented a 3D multi-stream U-Net, and proposed data prepossessing, that would normalise the data between MRI vendors. Six models were trained (two for each vendor, and two with combined vendor data) and their data was analysed, revealing that the combined vendor models, performed better than their singular data model for detecting the peripheral zone with a DSC of 0.81. However, it performed worse than the singular data models for detecting the whole prostate. This seems to be in line with the literature, where mixing datasets, can lead to poorer performance, than a model trained on a singular dataset \cite{zech2018variable,onofrey2019generalizable}. Liu et al. \cite{liu2020ms}, also worked on the problem of generalisation of segmentation models, implementing a Multi-site network that focused on learning the universal representation across the datasets and MRI scanners. They also developed a technique called Domain-Specific Batch normalisation, that was able to estimate the statistics and perform normalisation for each site the network had been trained on, helping to strengthen the segmentation. 

\section{Datasets}
\label{sec3}
The application of deep learning to the delineation of the prostate has achieved greater accuracy and results than previous manual or atlas-based methods \cite{onofrey2019generalizable}. To achieve this, networks are typically trained on large datasets that are annotated for a certain task such as imagenet \cite{deng2009imagenet}.
However, to gather such a large dataset for medical imaging is fraught with challenges. 

Even though terabytes of data is being created within the NHS, and is stored on secure local servers, there are governance issues, that restrict the sharing of this data between health centres and deep learning practitioners. The medical data, must be stored locally and only be accessed on site. This could be negated through the use of decentralised federated learning \cite{shi2019distributed}, where the model moves where it is training and the data stays at the local site. Which allows for privacy and security concerns to be addressed regarding. However, the infrastructure and logistics are currently not in place to allow this to be explored yet in practice \cite{onofrey2019generalizable}. Though this data may be available, it may not be labeled, or annotated in a way to allow it to be used with deep learning. To curate and collect medical imaging datasets, can be laborious, tedious and expensive to annotate new images. Thus it is important to explore current datasets and exploit them to create a generalisable network for delineating the prostate \citep{onofrey2019generalizable}. 

 \subsection{Limited Availability of Medical Imaging}

Within the United Kingdom, the National Health Service maintains the Diagnostic Imaging Dataset (DID) which has detailed information of diagnostics imaging performed but does not retain the actual medical images. Diagnostic Imaging Dataset (DID) \cite{nhschoices2020} From April 2019 to April 2020, a total of 42,330,840 imaging were performed. There is very limited publicly available imaging dataset in the United Kingdom. Only 3 large databases were identified as follow:

\begin{itemize}
    \item{\textbf{UK Biobank}. Largest UK biobank established for long-term follow up on 500,000 healthy volunteer. Dataset of abdominal, brain and heart MRI. Suitable for AI training and access on application.}
    \item{\textbf{REQUITE collaboration}. Largest comprehensive radiotherapy database on breast (n=2069), prostate (n=1808) and lung (n=561) cancer. Radiotherapy dataset includes radiotherapy images DICOMs (n=17,107) and DVHs (n=12684). Data on patient toxicities are also available. Excellent radiotherapy database for AI training with access on application\cite{seibold2019requite}.}
    \item{\textbf{National COVID-19 Chest Image Database (NCCID)}. Newly formed national database to collect X-ray, CT and MRI of COVID-19 patients. Data collection phase. Aimed for development and validation for AI machine learning. Data access on application.}
\end{itemize}

Limited good quality radiology and oncology imaging are available for artificial intelligence which will likely continue to be a barrier for artificial intelligence development for healthcare within the UK \cite{the_national_cancer_research_institute_2018}.
For the purpose of the segmentation of the prostate for this paper, we worked with 4 publicly available datasets. These included:

\begin{itemize}
    \item{\textbf{The Promise 12 dataset \cite{litjens2014evaluation}}. Which consists of 50 patients  T2 MRI scans of the prostate region. The data is this dataset, is representative of a clinical setting, and has been acquired from multiple centres, that use a variety of vendors and acquisition protocols. A Radiographist resident or image analysis researcher, created the contours for segmentation, with a resident radiographer with 6 years of MRI prostate experience checked and corrected the contours.}
    \item{\textbf{Prostate X \cite{litjens2017prostatex}}. The largest dataset of prostate data contains 300 patients data, in either T3 or T2 MRI scans. 98 T2 weighted MRI scans have  been annotated by a a medical student and then corrected by a urologist expert \cite{meyer2019towards}.}
    \item{\textbf{NCI-ISBI 2013 Challenge \cite{bloch2015nci}}. This dataset contains 60 patient Dicom files, of 1.5 MRI scans and T3 MRI SCANS. These were acquired from Radboud University Nijmegen Medical Centre [RUNMC],Netherland. The contours created for segmentation were either created by Drs. Nicolas Bloch (Boston University School of Medicine) and Mirabela Rusu (Case Western University) or Drs. Henkjan Huisman, Geert Litjens, or Futterer at RUNMC}
    \item{\textbf{The Decathlon Medical dataset \cite{simpson2019large}}. The decathlon dataset consists of 32  multi-modal MRI scans of the prostate, which has been acquired from RUNMC. The data was donated by Drs Geert Litjens and Bram van Ginneken from RUNMC.}
\end{itemize}

 As summarised in Table \ref{tab1}, a majority of previous research used private databases for their experiments, which makes it difficult to reproduce their work. This makes it difficult to ascertain if the modifications they have mentioned contributes to increase in their various metric scores or if the model is well tuned \cite{isensee2019automated}. 

The delineation of the prostate is a labour-intensive process performed manually by a trained clinician but is essential for radiotherapy. This is a time-consuming process, and the delineated prostate annotation can be subject to a variety of bias \cite{roques2014patient,njeh2008tumor,ghose2012survey}, but an accurate delineation is critical when planning treatments for patients. 

\begin{table*}
	\caption{\label{tab1}Summary of different methods and datasets pertaining to MRI prostate segmentation. DSC--Dice Similarity Coefficient, FPR--False Positive Rate, FNR--False Negative Rate, HD--Hausdorff Distance, AVD--Absolute Volume Distance, ABD--Absolute Boundary Distance, ASD--Average Surface Distance, RVD--Relative Volume Difference, aRVD--absolute Relative Volume Difference, MSD--Mean and Standard Deviation, AAD--Average Absolute Distance, MAD--Maximum Absolute Distance}%
	\centering
\begin{tabular}{|p{2cm}|p{4.5cm}|p{2cm}|p{2cm}|p{3cm}|}
			\hline
			Publication & Datasets & Metrics & Modality & Loss\newline Functions\\
			\hline
			\cite{tian2018psnet} & 41 patient in house dataset\newline 
			Promise12 \cite{litjens2014evaluation}\newline 
			ISBI2013 \cite{bloch2015nci}& DSC, RVD, HD, ASD& T2W MRI & Weighted Cross Entropy  \\ \hline
			\cite{clark2017fully} & 104 in house dataset\newline Promise12 \citep{litjens2014evaluation} & DSC\newline Sensitivity\newline	Specificity\newline	Accuracy \newline & mpMRI\newline T2W MRI & Dice loss  \\ \hline 
		 	\cite{karimi2018prostate} & 49 patient in house dataset \newline 26 training images Promise12 \cite{litjens2014evaluation}&DSC & T2W MRI & Squared Euclidean distance with elastic-net regularization\\ \hline
			\cite{zhu2017deeply} & 81 patient in house dataset & DSC & 3T MRI & Dice Loss \\ \hline
			\cite{milletari2016v} & Promise12 \cite{litjens2014evaluation}&  DSC, HD & T2 MRI & Dice Loss \\ \hline
			\cite{zhu2019fully} & 163 patients in house dataset& DSC, FPR, FNR & DWI, T2W MRI & Dice loss \\ \hline
			\cite{yu2017volumetric} & Promise12 \cite{litjens2014evaluation}& DSC, aRVD, ABD, HD& T2W MRI & Weighted Cross Entropy Loss\% \\ \hline
			\cite{wang2019deeply} &40 in house patients\newline Promise12 \cite{litjens2014evaluation} &  DSC, HD, aRVD, MSD& T2W MRI & Cosine Similarity and Cross Entropy \\ \hline
			\cite{liu2019automatic} & 319 patients from Prostate X \cite{litjens2017prostatex}\newline 46 patients in house dataset &  DSC & 3T MRI & Cross Entropy \\ 
			\hline
			\cite{zabihollahy2019automated} & 225 patients in house dataset&  DSC, AVD, Precision, Recall, Accuracy& 3T MRI & Dice Loss \\
			\hline
			\cite{nie2019semantic} & 50 patients in house dataset \newline Promise12\cite{litjens2014evaluation} &  DSC, ASD & T2W MRI & Soft Cross Entropy  \\
			\hline
			\cite{zhu2019boundary} & Promise12 \cite{litjens2014evaluation}\newline Phillips 3T MRI dataset\cite{rundo2017automated} \newline Brigham and Women’s Hospital Multiparametric MRI \cite{fedorov2018annotated} &  DSC, HD, ABD, RVD&T2W MRI, 3T MRI, mpMRI  & Boundary weighted Loss function\\
			\hline
			\cite{isensee2019automated} & Decathlon Prostate Dataset \cite{simpson2019large}\newline Promise12 \cite{litjens2014evaluation} & DSC & T2W MRI & Dice and Cross entropy loss \\
			\hline
			\cite{rundo2019use} & Philips Dataset\cite{rundo2017automated} \newline I2CVB\cite{lemaitre2015computer} \newline ISBI2013\cite{bloch2015nci} &  DSC, Sensitivity, Specificity, SPC, AAD, MAD & T2W MRI & Dice Loss \\
			\hline
			\cite{zavala2020segmentation} & 550 patient in house dataset &  DSC & T2W & Dice Loss \\
			\hline
			\cite{liu2020ms} & ISBI 2013\cite{bloch2015nci}\newline I2CVB\cite{lemaitre2015computer}
			& DSC,AVD & 3T MRI, T2W MRI & Dice Loss \\\hline
		
		\end{tabular}
	\end{table*}

\section{Comparative Study}
\label{sec5}
As part of this review, we implement a 2D U-Net architecture using the Ranger optimiser \cite{wright2019ranger} and Mish Activation \cite{misra2020mish}, to segment the prostate from MRI data. The ranger optimiser  is a combination of two optimisers, Rectified Adam \cite{liu2019variance} and Lookahead \cite{zhang2019lookahead} optimisers. These were inspired by recent work, in understanding the loss surfaces, and how initial training can have a large impact on the convergence of networks \cite{liu2019variance}. LookAhead keeps two sets of weights, and interpolates between them during training, it allows one set of faster weights to explore the loss landscape to choose an optimum path, where the slower weights stay behind to provide long term stability \cite{liu2019variance}. In the case of the Ranger optimiser, the faster weights use the Rectified Adam weights \citep{wright2019ranger}. Mish is a new activation function that can be displayed mathematically as 
\begin{equation}
x = x tanh(softplus(x))
\end{equation}
It is a self regularising  non-monotonic activation function, which has improved the performance of other networks, just by changing the activation functions within these networks to Mish \cite{wright2019ranger}.

We evaluate the ability of these slight changes to the U-Net configuration and its effect on performances on four publicly available datasets, i.e. the Promise 12 \cite{litjens2014evaluation}, Prostate X \cite{litjens2017prostatex}, NIC ISBI 2013 \cite{bloch2015nci} and the Decathlon Medical Dataset \cite{simpson2019large}. We evaluate models trained on each dataset and a combination of all, on a test set with DSC. The goal of this approach is to segue to a generalisable segmentation approach, that is both repeatable, open and usable by clinicians and oncologists. All of the images were pre-processed with an adaptive histogram and normalized during training. All of our code is available at https://github.com/AIEMMU/MRI\_Prostate. 

\subsection{Training}
For training, each model was trained with the ranger optimiser, with a learning rate of 1e-3 and a weight decay of 0.01, and using a flat + cosine anneal training curve with PyTorch. Each model is trained for 20 epochs, then fine-tuned and trained for a further 20 epochs with a batch size of 8 at a resolution of 448px x 448px. The loss function for the model is a combination of Dice Loss \cite{milletari2016v}, focal loss \cite{lin2017focal} and cross-entropy. 

For each model, the data was split at a random 80:20 split for training and testing. For validation, the training data was also a random 80:20 split. To avoid the model overfitting, data augmentation was used to to create new versions of the same image.  \cite{zhang2016understanding}.The augmentations using during training the neural network, included rotation, translation and vertical or horizontal flips \cite{dong2017automatic}. Additional augmentations were added to change the image quality using different contrast, lighting and adding noise.

The training was performed on a desktop computer with a 24GB RTX Titan, i5-9600K CPU, 32GB of RAM and a 2TB SSD drive. Training time for each model took on average 40 minutes, with overall time taking close to 6 hours for all 5 models. 


\subsection{Results}

\begin{table}[!ht]

  \small\addtolength{\tabcolsep}{2pt}
  \renewcommand{\arraystretch}{1.5}
\caption{The performance of the optimised 2D U-Net for MRI prostate segmentation on four datasets using Dice Similarity Coefficient (DSC).}
  \label{table:Results}
  \scalebox{0.85}{
    \begin{tabular}{cccccc}
        
        \multicolumn{6}{c}{Training Datasets}\\
        &Dataset  & Promise 12    & Prostate X  & Decathlon   & NIC ISBI            \\ \hline\hline
        \multirow{5}{*}{\rotatebox[origin=c]{90}{Testing Datasets}}
        &Promise12  & \textit{0.9069}     & 0.8869   & 0.9239   &0.828 \\
        &Prostate X  & 0.5891 &  \textit{0.9054}    & 0.9107   & 0.8011        \\
        &Decathlon&0.6055  & 0.6479  & \textit{0.9148} & 0.7395 \\
        &NIC ISBI  & 0.6931    & 0.7463   & 0.7628      & \textit{0.8909}         \\
        &Combined&\textbf{0.9171}  & \textbf{0.9061}  & \textbf{0.9427} & \textbf{0.8960} \\
        
        \hline
      \hline
  \end{tabular}}
\end{table}

Table \ref{table:Results} shows the obtained DSC scores on separate test sets for each dataset.  We observe that the model that was trained with all of the data outperformed all other models with higher DSC scores. However, we also observe that each of the networks performed well on the decathlon dataset apart from the NIC ISBI trained model. This could be due to a number of reasons. We show similar results to \cite{zavala2020segmentation}, that combining multiple datasets, resulted in an increase in DSC scores. 

\begin{figure*}[!ht]
    \centering
    \includegraphics[width=18cm]{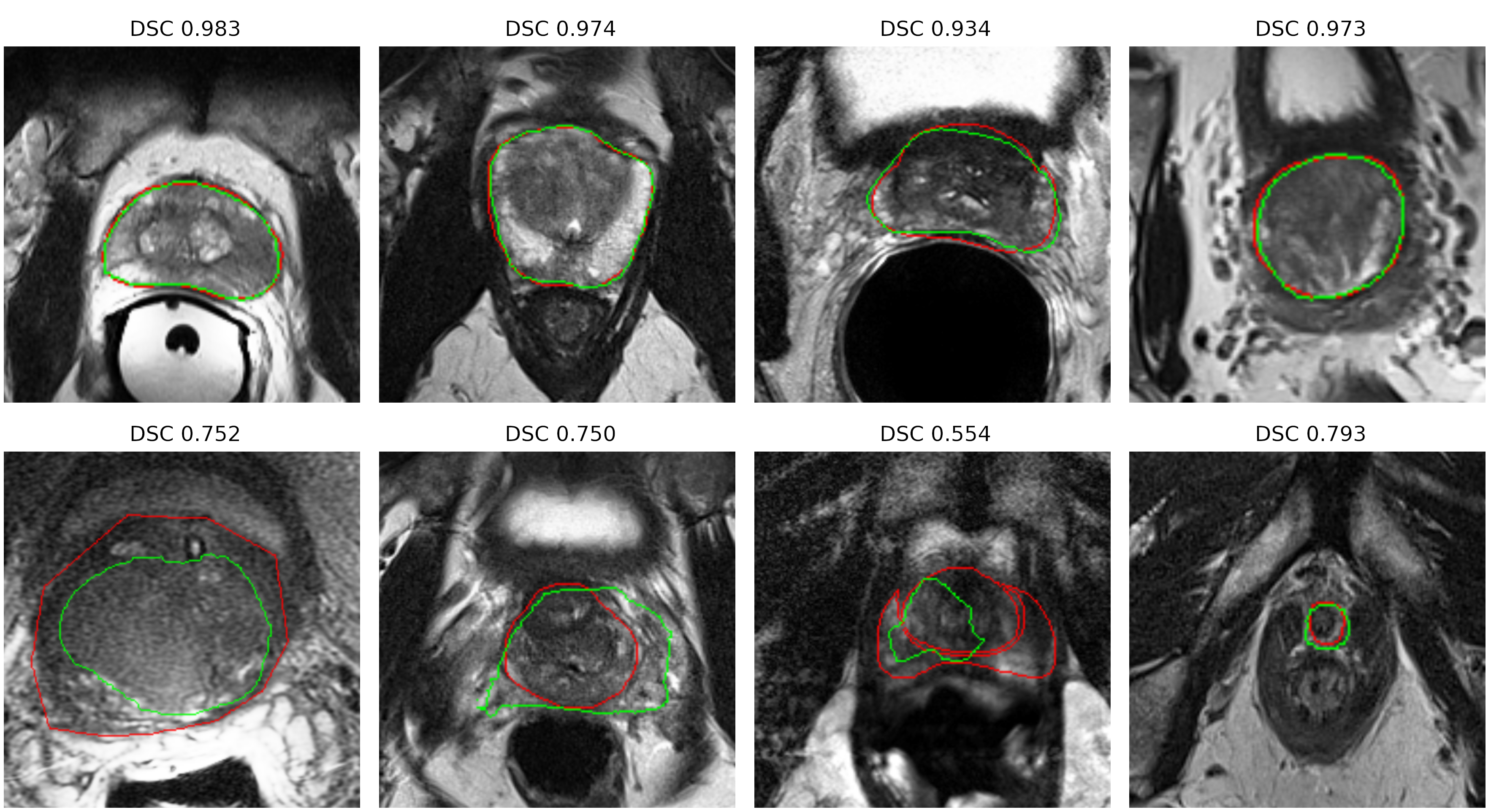}
    \caption[] {Examples of the auto prostate delineation, where green is the predicted outline and the red is the ground truth. The top row, are successes with DSC over 0.9. The bottom row are some challenging cases with poorer results. The images from left to right are from the Promise12 dataset, ProstateX dataset, NIC ISBI, Decathlon.}
    \label{fig:results}
\end{figure*}

Figure \ref{fig:results} presents the qualitative results of the segmentation, where the images are from separate Datasets. From left to right, the images are from: Promise12, ProstateX, NIC ISBI and decathlon. Thus, highlighting the capabilities of the network.

\section{Challenges, Recommendations and Future Research Direction}
This section addresses the challenges of deep learning in MRI prostate segmentation. We provide some recommendations for future research direction and draw a new perspective for this research domain. 

\subsection{Limited Annotated Data}
This paper has presented challenges that still need to be met within the field of segmentation of the prostate from MRI data. This includes the use of private data and a lack of publicly annotated datasets. As many papers within this study, use a private in house dataset, it is not possible to compare fairly against other networks and know what changes to a network impact the performance of a network as discussed in \cite{isensee2019automated}. Deep learning greatly improves their accuracy by being trained on large datasets within their domain. However, this is a difficult task within the medical imaging, especially with restrictions on patients data and the laborious task of collecting and annotating new images. This restricts the use of AI within a clinical setting. Although, it allows for researchers to innovate and explore new techniques to exploit the small datasets that are available using techniques. 

One of such is Mixup  a newer data augmentation technique \cite{zhang2017mixup}, which has improved results dramatically for small datasets, and for models that are not pretrained on similar data. Mixup is a simple concept, which combines two images via an alpha overlay. It has been shown to improve the generalisation of knee segmentation \cite{panfilov2019improving} for MRI data. This technique has been expanded to combine latent feature from the layers of a network \cite{verma2019manifold}. 

Another approach is using self-supervised learning, where the model can learn representations of the data without using annotated data. This can be accomplished by training the model on tasks such as recognising color spaces \cite{goyal2019scaling}, inpainting \cite{pathak2016context} or predicting image rotations. The learned features using self-supervised tasks are good alternatives for tasks with little data. Goyal et al. \cite{goyal2019scaling}, who via scaling the amount of data available for training, model parameters and hardness of the self-learning task, showed that they equal too or surpassed supervised learning tasks.  

\subsection{Work Flows}

Deep learning attempts to replicate the ground truth labels for segmentation, but variation between clinicians is recognised as one of the weakest link in the contouring process among cancer clinicians \cite{vinod2016uncertainties}. Clinicians do adopt a QA and peer review to reduce inter-observer variation \cite{huo2017evidence}. However, many clinicians rely on their own experiences and data to outline their own ground truths, which a deep learning model will not have access to. Thus it is important to rethink how segmentation models may be deployed in a clinical setting. Will they be replacing the annotation stage or offer a suggested contour that the clinician may use/edit or allow the clinician to create their annotation. 

\subsection{Performance Metrics}

The training of auto segmenting models via deep learning relies on metrics such as the DSC, HD, or class accuracy. However, these metrics are not sensitive to significant clinical errors. While DSC scores how similar images are, it is insensitive to small changes in border outlines, these small changes may have a significant impact on the deliberating effects for patients. Small offsets, may require new contours to be annotated completely, reducing any time that the deep learning system may have saved. Surface Dice has shown promise as a metric for measuring the performance of an auto segmenting model \cite{nikolov2018deep}. Yet, \cite{nikolov2018deep} to be able to use this metric, they had to employ 3 oncologists to annotate data, and measure an acceptable tolerance for differing organs. This limits the use of this metric, outside head and neck radiography. More metrics need to be explored, that can generalise well to differing organs, and be clinically acceptable as well. Future work will focus on this. 

Additionally, apart from using a standard performance metrics, for fair comparison, we recommend to establish standardised experiment protocols where the researchers can compare their algorithms on a standard dataset from multiple vendors. With standardisation of training set and testing set, the protocols can be used to focus solely on the design of the network and augmentation strategies. For reproducibility and transparency of the implementation, the participants would share their codes with full details of implementation, as in \cite{fastai} and our proposed method here as exemplar. 

\section{Conclusion}
In this paper, we have discussed state-of-the-art deep learning algorithms for MRI prostate segmentation. We have shown the challenges in conducting a fair comparison on the performance of these networks, as most are not reproducible due to their use of in-house datasets. We have demonstrated that an optimised 2D U-Net using existing activation function and optimisers, can achieve high DSC scores on 4 different test datasets. We have also demonstrated that networks trained on the Promise12 \cite{litjens2014evaluation} and ProstateX \cite{litjens2017prostatex} achieved high DSC scores on the Decathlon dataset \cite{simpson2019large} and generalise well to the other datasets. Finally, we address the limitations of MRI prostate segmentation and provide a new perspective. Future work will investigate metrics that can assess a models performance more in line with oncologists expertise.

\section*{Acknowledgments}
This project is funded by the Margaret Baecker Endowment Fund and the Clatterbridge Cancer Centre NHS Trust.

\bibliographystyle{model2-names.bst} \biboptions{authoryear}
\bibliography{prostate}

\section*{Supplementary Material}

The codebase for this paper, experiments and processing code is available at https://github.com/AIEMMU/MRI\_Prostate

\end{document}